# EIGENRAYS IN 3D HETEROGENEOUS ANISOTROPIC MEDIA:

# PART IV – GEOMETRIC SPREADING FROM TRAVELTIME HESSIAN

*Igor Ravve (corresponding author) and Zvi Koren, Emerson*

Igor.ravve@emerson.com , zvi.koren@emerson.com

## ABSTRACT

Considering 3D heterogeneous general anisotropic media, in Part III of this study we obtain the finite-element solution for the stationary ray path between two fixed endpoints using an original finite element approach. It involves computation of the global (all-node) traveltime gradient vector and Hessian matrix, with respect to the nodal location and direction components. Both the traveltime gradient and Hessian are used in the optimization process, applying the Newton method for searching minimum traveltime ray paths, and the gradient method for searching saddle-point stationary rays (which are an evidence of caustics). The global traveltime Hessian of the resolved stationary ray also plays an important role for obtaining the dynamic characteristics along the ray. It is a band matrix that includes the spatial, directional and mixed second derivatives of the traveltime at all nodes and nodal pairs and hence provides information about paraxial rays and local curvatures of the propagated wavefronts. In Parts V, VI and VII of this study we explicitly use the global traveltime Hessian for solving the dynamic ray tracing equation, to obtain the geometric spreading at each point along the ray. Moreover, the dynamic ray tracing makes it possible to identify and classify caustics along the ray.

In this part (Part IV) we propose an original two-stage approach for computing the relative geometric spreading of the entire ray path without explicitly performing the dynamic ray tracing. The first stage consists of an efficient algorithm for reducing the already computed global traveltime Hessian into a $6 \times 6$ endpoint spatial traveltime Hessian. The second stage involves application of a known workflow for computing the geometrical spreading using the $3 \times 3$ off-diagonal sub-block of the endpoint traveltime Hessian. Note, that the method proposed in this part (Part IV) doesn't deliver information about caustics, nor the geometric spreading at intermediate points of the stationary ray path. Throughout the examples presented in Part VII we demonstrate the accuracy of the method proposed in this part (Part IV) over a set of isotropic and anisotropic examples by comparing the geometric spreading computed from this method based on the traveltime Hessian reduction with the one computed from the actual dynamic ray tracing.

## INTRODUCTION

The method presented in this part (Part IV) can be considered a condense dynamic solution that effectively delivers a limited information without the need to perform the dynamic ray tracing. We use the global traveltime Hessian of the stationary ray path, with respect to (wrt) the nodal locations and directions, to compute the relative geometric spreading of the entire ray. The method consists of two stages. The first stage involves an original approach to condense the global traveltime Hessian into a $6 \times 6$ endpoint (source-receiver) traveltime Hessian. In the second stage we use its off-diagonal (mixed-derivative) $3 \times 3$ sub-block to compute the geometric spreading of the entire ray path, applying a known technique.

### Geometric spreading based on the endpoint traveltime Hessian: A review

As a signal moves away from a source, its wavefront normally (but not necessarily) expands (diverges) and the corresponding amplitude of the particle motion (or the energy flux density associated with the ray tube) decreases. Moreover, due to possible local focusing effects of the wavefront at some points along the ray, referred to as caustics, the ray amplitude may increase and the phase of the wavefront can change as well. The dynamic phenomena of amplitude and phase changes along waves/rays in elastic media are associated with the so-called complex-valued geometric spreading, or real-valued geometric spreading and phase-rotation factors (e.g., Sommerfeld, 1964, Červený, 2000; Chapman, 2004; Schleicher et al., 2007; Tsvankin and Grechka, 2011, and many others).

Amplitude decay of seismic waves has been attracting the attention of seismologists for over a century (e.g., Gutenberg, 1936). O'Doherty and Anstey (1971) analyzed the factors affecting true-amplitude seismic processing, and demonstrated that among the many amplitude decay factors, geometric spreading is the most dominant. Basing on the Gutenberg (1936) results for amplitudes of reflected waves, Newman (1973) studied the divergence effects (decrease of wave energy per unit area of the wavefront) vs. offset and depth, and obtained the divergence coefficients affecting the reflection magnitudes in terms of effective normal moveout velocities of multi-layer models. The above-mentioned dynamic characteristics are important for both seismic forward modeling and seismic migration/inversion. In particular, they are applied as the seismic modeling/migration amplitude weighting and phase correction components in the Kirchhoff (or Born) integral to obtain (data-domain) synthetic seismic seismograms and (image-domain) "true" amplitude subsurface seismic reflectivities (e.g., Sommerfeld, 1964, Beylkin, 1985; Bleistein, 1987; Najmi, 1996; Bleistein et al., 2001; Chapman, 2004; Schleicher et al., 2001, 2007; Koren and Ravve, 2011, and many others). This is obviously important for amplitude vs. offset/angle (AVO/AVA) analyses (e.g., Tygel et al., 1992; Schleicher et al., 1993; Vanelle and Gajewski, 2003; Xu and Tsvankin, 2008; Skopintseva et al., 2012; Tanaka et al., 2017, among others). There is a great wealth of published material available on the computation of dynamic properties along rays. Obviously, the following review is not complete, covering only a part of this comprehensive subject. As a principle reference, we refer to the comprehensive book of Červený (2000), with many references therein. Overall, geometric spreading can be obtained by two main approaches. The comprehensive method, described in Part V of this study, is based on solving the DRT equation (paraxial ray theory) in ray-centered or Cartesian coordinates (e.g., Virieux et al., 1988; Červený, 2001; Iversen et al., 2018, among others). This method provides computation of general paraxial rays and computation of a transformation matrix (e.g., the transformation matrix $\mathbf{Q}$ between the Cartesian or ray-centred coordinates and the ray coordinates at the source) along the ray path, which makes it possible to compute the geometrical spreading at any point along the ray and to identify caustic locations and classify their types. This part (Part IV) is devoted to the computation of the geometric spreading between the endpoints of the entire ray by directly using the second derivatives of the traveltime wrt the source and receiver location components (e.g., Gajewski and Pšenčík, 1987, Červený, 2000; Xu and Tsvankin, 2006). Below we review the published works on this topic.

Geometric spreading based on the traveltime Hessian wrt source/receiver locations

In many practical applications, only the geometric spreading between a source and receiver is needed, ignoring phase shifts due to caustics. In this case, dynamic ray tracing is not a must, and geometric spreading can be obtained from the mixed (off-diagonal) block of the source-receiver traveltime Hessian matrix. This symmetric endpoint matrix of dimension $6\times6$ includes the external degrees of freedom (DoF) of the source, $S$, and receiver, $R$, locations only. It consists of four blocks of dimension $3\times3$, where the upper and lower symmetric diagonal blocks are related to the source and receiver, respectively, while the two off-diagonal blocks, transposed to each other, are related to both the source and receiver. Goldin (1986) suggested a workflow to compute the geometric spreading using the $3\times3$ off-diagonal block, with its last line and column removed (related to the coordinate axis normal to the acquisition surface), so that the resulting "mixed" Hessian matrix $\mathbf{M}_{RS}$ becomes $2\times2$.

Hron et al. (1986) analyzed geometric spreading in layered isotropic media, where the propagation becomes effectively 2D. In this case, the matrix of the mixed lateral spatial derivatives $\mathbf{M}_{RS}$ of the traveltime $t$ (of dimension $2\times2$),

$$\mathbf{M}_{RS} = \frac{\partial^2 t}{\partial x_{R,i} \partial x_{S,j}} \quad , \quad i,\ j = 1,\ 2 \quad , \tag{1}$$

becomes diagonal and its determinant can be factorized,

$$\det \mathbf{M}_{RS} = L^{||} L^{\perp} \quad , \tag{2}$$

where $R$ and $S$ are labels related to the source and receiver, while $L^{||}$ and $L^{\perp}$ are in-plane and out-of-plane factors, respectively,

$$L^{||} = \frac{d^2 t}{dh^2} \quad , \quad L^{\perp} = \frac{1}{h} \frac{dt}{dh} \quad , \tag{3}$$

and $t(h)$ is the traveltime at the offset $h$. The isotropic geometric spreading $L_{GS}$ becomes,

$$L_{GS} = \sqrt{\frac{\cos\theta_R \cos\theta_S}{\det \mathbf{M}_{RS}}} \quad , \tag{4}$$

where $\theta_S$ and $\theta_R$ are take-off angles at the source and receiver points, respectively. The absolute value of the expression under the square root is used in case it becomes negative.

Ursin (1990) applied these expressions to derive a closed form approximation formula of geometric spreading for traveltime $t(h)$ in the forms of a) a simple hyperbolic moveout approximation, and b) a shifted hyperbola moveout (Castle, 1988).

Najmi (1996) obtained closed form solutions for the geometric spreading of an acoustic wave in isotropic inhomogeneous media for both flat and dipping reflectors, considering traveltime functions and their spatial derivatives wrt endpoint locations. Schleicher et al. (2001) extended the Kirchhoff–Helmholtz integral to general anisotropic media and obtained exact expressions for the geometric spreading in terms of phase and ray velocities. Vanelle and Gajewski (2002) suggested using hyperbolic traveltime approximation in order to interpolate the traveltime in the vicinities of the source and receiver, in order to compute geometric spreading. In their later studies (Vanelle and Gajewski, 2003, 2013; Vanelle et al., 2006), equation 4 for geometric spreading was refined,

$$L_{GS} = \sqrt{\cos\theta_{\mathrm{ray},R} \cos\theta_{\mathrm{ray},S} \frac{v_{\mathrm{ray},S}}{v_{\mathrm{phs},S}} \frac{v_{\mathrm{ray},R}}{v_{\mathrm{phs},R}} \frac{1}{\det \mathbf{M}_{RS}}} \quad , \tag{5}$$

accounting for the transform between the Cartesian coordinates and the RCC in anisotropic media (Klimeš, 2006), or alternatively,

$$L_{GS} = \sqrt{\frac{\cos\theta_{\mathrm{ray},S}}{\cos\beta_S} \frac{\cos\theta_{\mathrm{ray},R}}{\cos\beta_R} \frac{1}{\det \mathbf{M}_{RS}}} \quad , \tag{6}$$

where $v_{\mathrm{phs},S}$, $v_{\mathrm{phs},R}$ and $v_{\mathrm{ray},S}$, $v_{\mathrm{ray},R}$ are the phase and ray velocities, respectively, $\theta_{\mathrm{ray},S}$ and $\theta_{\mathrm{ray},R}$, are the ray (group) take-off angles at the endpoints of the path (i.e., angles between the ray velocities and the normal directions to the acquisition surfaces), $\beta_S$ and $\beta_R$ are angles between the phase and ray velocities, and subscripts $S$ and $R$ are related to the source and receiver, respectively.

Ursin and Hokstad (2003) studied layered azimuthally isotropic (e.g., VTI) media and concluded that in this case the determinant of the mixed matrix of the traveltime Hessian $\mathbf{M}_{RS}$ has the same expression as for isotropic layered media, i.e., equations 2 and 3 hold.

Mirko van der Baan (2004) applied plane-wave decomposition and studied the moveout correction in the $\tau - p$ domain, where the waves are planes, unlike the spherical waves in the $t - h$ domain. In homogeneous media, both isotropic and anisotropic plane waves are not distorted with the increase of the propagating distance, and are not subject to geometric spreading. This statement remains valid for horizontally-layered media with constant layer parameters.

Xu et al. (2005) studied geometric spreading for P-waves in laterally homogeneous horizontally-layered triclinic media. For such media, the geometric spreading becomes a function of reflection traveltime and its spatial derivatives (Tsvankin and Grechka, 2011),

$$
\begin{aligned}
\det \mathbf{M}_{RS} = {} & \frac{\partial^2 t}{\partial h^2}\frac{\partial t}{\partial h}\frac{1}{h} + \frac{\partial^2 t}{\partial h^2}\frac{\partial^2 t}{\partial \psi_{\text{off}}^2}\frac{1}{h^2} - \left(\frac{\partial t}{\partial \psi_{\text{off}}}\right)^2 \frac{1}{h^4} \\
& - \left(\frac{\partial^2 t}{\partial h \partial \psi_{\text{off}}}\right)^2 \frac{1}{h^2} + 2\frac{\partial^2 t}{\partial h \partial \psi_{\text{off}}}\frac{\partial t}{\partial \psi_{\text{off}}}\frac{1}{h^3} \quad ,
\end{aligned}
\tag{7}
$$

where $t\left(h, \psi_{\text{off}}\right)$ is the traveltime, and $h$ and $\psi_{\text{off}}$ are the offset magnitude and azimuth, respectively. For monoclinic and orthorhombic layered media, the equation remains the same; in the vertical planes of symmetry of orthorhombic media, the odd azimuthal derivatives vanish; for VTI media all azimuthal derivatives vanish (Tsvankin and Grechka, 2011). Note that equations 2 and 3 are a particular case of equation 7; these equations are exact. Equation 7 can be combined with any of the known high-order (multi-parameter) traveltime approximations published in the literature, such as Tsvankin and Thomsen (1994) four-parameter or Alkhalifah and Tsvankin (1995) three-parameter moveout approximations. Both approximations include an asymptotic correction in the denominator of the nonhyperbolic term. The two moveout approximations were initially suggested for azimuthally isotropic layered media, but were later extended and used for azimuthally anisotropic media by Vasconcelos and Tsvankin (2006). Xu et al. (2005) combined equation 7 with the four-parameter approximation, and Xu and Tsvankin (2006) combined it with the three-parameter approximation. In both cases, the azimuthally dependent geometric spreading approximations were obtained for compressional waves in layered orthorhombic media. In a later study, Xu and Tsvankin (2008) applied the four-parameter moveout approximation to obtain the geometric spreading for converted PSV waves in layered VTI media.

Stovas and Ursin (2009) derived an approximation for geometric spreading for compressional waves in transversely isotropic layered media with vertical axis of symmetry (VTI) vs. three effective parameters: normal incidence time, NMO velocity, and the heterogeneity coefficient (the latter can be replaced by the effective anellipticity). Golikov and Stovas (2013) derived a moveout-based geometric spreading approximation for transversely isotropic media with tilted axis of symmetry (TTI). Stovas (2017, 2018) and Shibo et al. (2018) derived analytical formulae to compute geometric spreading for a stack of acoustic orthorhombic layers with azimuthal variations of their symmetry planes. Ivanov and Stovas (2017) derived a mapping operator that allowed estimation of the geometric spreading in tilted (rotated) anisotropic symmetries from the geometric spreading in their 'crystal' frames, and demonstrated the accuracy and efficiency of the method for a tilted orthorhombic medium. Xu et al. (2018) developed geometric spreading approximations for layered orthorhombic media vs. normal incidence time, high and low effective NMO velocities, and three effective anellipticities, where VTI can be considered as a particular case with a single effective NMO velocity and a single effective anellipticity. Recently, Xu and Stovas (2018) derived a geometric spreading formula based on the generalized moveout approximation suggested by Fomel and Stovas (2010).

In the abstracts by Koren and Ravve (2019) and Ravve and Koren (2019), we show that the global traveltime Hessian $\nabla_{\mathbf{d}}\nabla_{\mathbf{d}} t$, that includes the external and internal DoF: location and direction components at all nodes, $\mathbf{d} = \{\mathbf{x}, \mathbf{r}\}$, can be reduced to a $6\times6$ source-receiver traveltime Hessian matrix by eliminating all the internal DoF (other than the source and receiver locations). Its off-diagonal $3\times3$ block is further reduced to the above mentioned $2\times2$ matrix $\mathbf{M}_{RS}$, which can then be used to compute the geometric spreading. In this paper, we explain in details the method with all the required derivations.

Note that the global traveltime Hessian $\nabla_{\mathbf{d}}\nabla_{\mathbf{d}} t$ is positive-definite (with all eigenvalues positive) when the stationary path delivers a minimum traveltime, and indefinite (with eigenvalues of both signs) when it yields a saddle point traveltime. The former case corresponds to a caustic-free path, and the latter to a path with caustics. This method, however, cannot reveal the number of caustics (if any), their location along the path and their types (orders).

## GENRAL REMARKS

Types of stationary rays and existence of caustics

Assuming the stationary ray path has already been found, we first analyze whether the path delivers a minimum traveltime solution by computing the eigenvalues of the already computed, global traveltime Hessian matrix. If all the eigenvalues are positive, the stationary path yields a minimum traveltime. We assume that the (hypothetic) case where all the eigenvalues are negative, which means a maximum traveltime solution, is unlikely to happen for systems with multiple degrees of freedom (DoF). If one or more eigenvalues are negative, and the others are positive, it is a saddle point solution, which is an indication for the existence of caustics: zeros of the (signed) cross-section area of the ray tube (ray Jacobian) along the ray. However, this type of analysis cannot provide the number of caustics, their locations and their type: line or point, which are important for implementing the required phase correction of the Green's function.

Geometric spreading without dynamic ray tracing

Geometric spreading represents the amplitude (normally decay) factor of the Green's function. In this study, the term "geometric spreading" refers to the so-called (reciprocal) relative geometric spreading (e.g., Červený, 2000). In cases where we are only interested in the total geometric spreading between the endpoints, we propose an original and efficient approach for computing the geometric spreading without explicitly solving the dynamic ray tracing (DRT) equations. The method is based on reducing the global traveltime Hessian into a $6\times6$ endpoint spatial traveltime Hessian, where only the off-diagonal sub-matrix (with the "mixed" source-receiver spatial derivatives) is needed. The term "mixed" means that one Cartesian coordinate in the second derivative belongs to the source and the other to the receiver. The method works for ray paths with caustics as well.

Degrees of Freedom

The independent parameters of the traveltime function are only the locations of the endpoints of the ray path (source and receiver coordinates), refer to as the external DoF. The locations and ray directions at the internal nodes and the directions at the endpoints are dependent parameters defined by the endpoints' location, medium properties, stationarity condition and the nodal density distribution constraint, refer to as internal DoF. Given the source-receiver locations (external DoF), the internal DoF can be all computed. Furthermore, for infinitesimal variations of the external DoF, the change of the internal DoF is also infinitesimal, and given the global (all-node) traveltime Hessian, this dependence can be linearized. An infinitesimal change of the external DoF yields a paraxial ray. Thus, we compute the change of the traveltime (as compared to that of the central ray) vs. the change of the source-receiver coordinates only, and we then establish the Hessian of this traveltime function. The symmetric $6\times6$ endpoint traveltime Hessian consists of four $3\times3$ blocks. The upper and lower symmetric diagonal blocks are related to the source and receiver, respectively. The two mixed off-diagonal blocks are transposed to each other and related to both the source and receiver. Only a mixed block is needed to compute the geometric spreading. The orientation of the acquisition surface is taken into account, and then the mixed block is further reduced from dimension $3\times3$ to $2\times2$ whose determinant is a key component of the relative geometric spreading.

Appendices

Appendix A includes the original derivations needed to condense the global traveltime Hessian into the source-receiver $6\times6$ Hessian.

Appendix B includes a known conventional background material needed to compute the relative geometric spreading from the mixed $3\times3$ block of the endpoint (source-receiver) traveltime Hessian.

## GEOMETRIC SPREADING WITHOUT DYNAMIC RAY TRACING

The Eigenray method suggested in this study provides a natural way to compute the geometric spreading for the whole ray path between the source and receiver without explicitly performing dynamic ray tracing (DRT). The DRT solution, described in Parts V and VI and implemented in Part VII, is required in order to compute and analyze the ray Jacobian, $J(s)$ (signed cross-area of the ray tube), for detecting possible caustic locations and for computing the geometric spreading, $L_{GS}(s)$, for any point along the ray path.

Along with the ray path, arclength and traveltime, the Eigenray method provides the global traveltime Hessian matrix of the stationary path. The latter represents a large, narrow-band, square matrix that includes all DoF before imposing the boundary conditions. In this part of our study, we suggest a way to compress this matrix into a $6 \times 6$ endpoint traveltime Hessian, whose DoF are spatial locations of the source and receiver, $\mathbf{x}_S$ and $\mathbf{x}_R$. In this section and in Appendix A, we provide the full theory briefly presented in the abstract by Koren and Ravve (2019).

The geometric spreading is obtained directly from the compressed Hessian matrix of the traveltime wrt the endpoint locations. Each endpoint has three spatial DoF (coordinates); therefore, the endpoint traveltime Hessian is a symmetric matrix of dimension $6 \times 6$. It consists of four $3 \times 3$ blocks. The upper diagonal block is related to the source, the lower diagonal block to the receiver, and the two off-diagonal "mixed derivative" blocks are related to both source and receiver. The diagonal blocks are symmetric, while the off-diagonal blocks are the transpose of each other. Unlike the global traveltime Hessian that includes all nodal DoF and is subject to the boundary conditions, the endpoint traveltime Hessian is not necessarily positive definite even when the stationary path delivers the minimum traveltime.

The geometric spreading is computed in two stages:

- 'Condensing' the global traveltime Hessian matrix $\nabla_{\mathbf{d}} \nabla_{\mathbf{d}} t$ to the endpoint traveltime Hessian matrix $\widetilde{\nabla_{SR} \nabla_{SR}} t$, where subscript d means all DoF (related to locations and directions), while $S$ and $R$ are related to the source and receiver locations, respectively. The Hessian $\nabla_{\mathbf{d}}\nabla_{\mathbf{d}}\, t$ is a narrow-band square matrix of length $6 \times$ number of nodes. Its band width is 12 for two-node elements and 18 for three-node elements. The dimension of Hessian $\widetilde{\nabla_{SR}\nabla_{SR}}t$ is $6\times6$ and we extract its $3 \times 3$ off-diagonal mixed block.

- Applying a conventional (well-known) method for computing the geometric spreading, given the extracted $3 \times 3$ mixed Hessian and other computed data.

We emphasize that when computing the Eigenray stationary path, the location components of the endpoints are fixed and not considered DoF. The DoF are the location and direction components of the internal nodes and also the direction components of the end nodes. On the other hand, when computing the geometric spreading for the whole ray, only the location components of the two end nodes (a total of six components) are considered independent DoF. The locations of the internal nodes and the directions of the ray at all the nodes (including the end nodes) are internal, dependent DoF. Given the locations of the two end nodes, all these internal DoF are fully defined by the stationary ray. There may be several stationary ray paths for the same endpoint locations (multi-pathing); each of them is treated independently when computing the geometric spreading.

In the Eigenray approach, the traveltime depends on all DoF, $\mathbf{d}$, where $t = t\left(\mathbf{d}_S, \mathbf{d}_A, \mathbf{d}_R, \mathbf{d}_B\right)$, where $A$ and $B$ are indices of internal DoF. Specifically, $B$ stands for the direction DoF of the receiver, and $A$ for all other DoF (direction DoF of the source and location and direction DoF of all internal nodes of the path). However, if the external DoF (endpoint locations) are fixed, then the internal DoF, $\mathbf{d}_A = \mathbf{d}_A\left(\mathbf{d}_S, \mathbf{d}_R\right)$ and $\mathbf{d}_B = \mathbf{d}_B\left(\mathbf{d}_S, \mathbf{d}_R\right)$, can be explicitly computed, accounting for the vanishing traveltime gradient. Hence, the internal DoF can be excluded and the traveltime equation becomes only a function of the source and receiver locations,

$$t = t\left[\mathbf{d}_S, \mathbf{d}_A\left(\mathbf{d}_S, \mathbf{d}_R\right), \mathbf{d}_R, \mathbf{d}_B\left(\mathbf{d}_S, \mathbf{d}_R\right)\right] = t\left(\mathbf{d}_S, \mathbf{d}_R\right) \quad . \tag{8}$$

The corresponding 6 × 6 endpoint traveltime Hessian exists and reads,

$$\nabla_{SR}\nabla_{SR}\, t = \begin{bmatrix} \nabla_S\nabla_S\, t & \nabla_S\nabla_R\, t \\ \nabla_R\nabla_S\, t & \nabla_R\nabla_R\, t \end{bmatrix} \quad . \tag{9}$$

The tilde symbol is needed to distinguish these blocks from the corresponding blocks of the global traveltime Hessian. Each of the four blocks in the matrix of equation 9 is 3 × 3. The implementation details for computing the traveltime Hessian matrix of equation 9 are explained in Appendix A. The algorithm includes inversion of the "stiffness" submatrix related to the internal DoF. However, this submatrix becomes invertible (i.e., its determinant does not vanish) only after implementing the constraints which are explained below. The physical reason for the vanishing determinant is that the nodes of a stationary paraxial ray can move along its path without actually affecting the path geometry (i.e., the paraxial ray path remains the same). Thus, the stationarity condition does not fully define the location of the nodes of the paraxial ray. A similar indefiniteness exists for the central ray – that is why we use the node density distribution constraint for the central ray. To define the nodal locations of the paraxial ray, we assume that the shifts between the nodes of the central ray and paraxial rays are normal to the central ray. The left lower 3 × 3 block consisting of the mixed derivatives (i.e., those derivatives where one coordinate belongs to the source and the other to the receiver) is needed to apply a conventional well-known approach for computing the geometric spreading (Goldin, 1986; Červený, 2000; Tsvankin and Grechka, 2011, among others), as explained in Appendix B. As mentioned, the geometric spreading is a function of the compressed endpoint traveltime Hessian which, in turn, is a function of the global traveltime Hessian, related to the stationary ray path.

We have already noted that the (relative) geometric spreading is insensitive to the swapping of the source and receiver. It does not matter whether it has been computed by applying the endpoint traveltime Hessian or by using the ray Jacobian (dynamic ray tracing, as we propose in Parts V, VI and VII).

The compressed endpoint traveltime Hessian is a $6\times6$ matrix that includes the following four $3\times3$ blocks,

| | |
|---|---|
| $SS$ | $SR$ |
| $RS$ | $RR$ |

where label $S$ means three DoF characterizing the location of the source, and label $R$ means those of the receiver. The $6\times6$ endpoint traveltime Hessian is symmetric. It consists of two diagonal symmetric blocks $SS$ and $RR$ related to the source and receiver, respectively, and two mixed blocks related to both the source and receiver, which are transposed to each other, $RS^T = SR$. Now assume that the source and receiver have been swapped. The diagonal blocks exchange their places with each other, and the mixed blocks are replaced by their transpose. After swapping the source and receiver, the endpoint traveltime Hessian looks like,

| | |
|---|---|
| $RR$ | $RS$ |
| $SR$ | $SS$ |

However, to compute the geometric spreading, we need only the determinants of the mixed block of dimension $3\times3$ (to be further reduced to dimension $2\times2$). The transposition operator does not affect this determinant. Appendix B provides all the required details for computing the geometric spreading with the use of the mixed block $RS$ of the endpoint traveltime Hessian.

Ray path complexity criterion

In order to estimate the reliability/plausibility of the ray path obtained with the Eigenray KRT, we introduce the *endpoint* propagation complexity criterion,

$$\hat{c}_r = \left(L_{GS} / \sigma - 1\right)^2 \quad . \tag{10}$$

A zero complexity corresponds to an isotropic medium with a constant (not necessarily vertical) velocity gradient, where the relative geometric spreading coincides with parameter sigma, defined as $d\sigma = v_{\text{ray}} ds = v_{\text{ray}}^2 d\tau$, i.e., $L_{GS} = \sigma$, and thus, the endpoint propagation complexity $\hat{c}_r$ vanishes. In part V we define the *weighted* propagation complexity $c_r \neq \hat{c}_r$ that accounts for the normalized relative geometric spreading not only at the receiver point (implemented in equation 10), but along the whole stationary ray path.

## NUMERICAL EXAMPLES

The numerical examples of the proposed method are presented in Part VII, where for a set of isotropic and anisotropic models, we compare the geometrical spreading values obtained by the proposed method based on the endpoint traveltime Hessian with those obtained by the dynamic ray tracing. We demonstrate that for the entire ray path, the results of the two methods coincide with a negligible computational error.

## CONCLUSIONS

We propose an original and efficient algorithm for computing the geometrical spreading of the entire stationary ray, applying the global (all-node) traveltime Hessian, already computed along the resolved stationary ray path. We first reduce (compress) the global traveltime Hessian to the endpoint (source-receiver) spatial traveltime Hessian, which makes it possible to apply the known technique using its off-diagonal mixed-derivative sub-block. The proposed method is very efficient; however, it doesn't explicitly deliver the important information about possible caustics along the ray, their location and their type (point or line). Moreover, it does not deliver the geometric spreading between the source and intermediate points along the ray path. Nevertheless, there may be still a class of problems in which the information only about the total geometric spreading, provided in this study, is valuable.

## ACKNOWLEDGEMENT

The authors are grateful to Emerson for the financial and technical support of this study and for the permission to publish its results. The gratitude is extended to Ivan Pšenčík, Einar Iversen, Michael Slawinski, Alexey Stovas, Vladimir Grechka, and our colleague Beth Orshalimy, whose valuable remarks helped to improve the content and style of this paper.

## APPENDIX A. REDUCTION OF GLOBAL HESSIAN TO ENDPOINT HESSIAN

In this appendix, we compress the full narrow-band square Hessian matrix of dimension $6 \times (N+1)$, that includes all DoF for $N+1$ ray path nodes, to the $6 \times 6$ endpoint traveltime Hessian that only includes spatial DoF of the source and receiver. This endpoint traveltime Hessian matrix is first analyzed to verify whether it is positive definite (corresponding to a minimum traveltime solution) or indefinite (corresponding to caustic locations along the ray). We then use it for computing the geometric spreading between the endpoints of the ray.

For a stationary path, the gradient of the target function that includes the traveltime and the weighted penalty terms, vanishes. In the analysis and computation of the geometric spreading, we can ignore the penalty terms, assuming that the contribution of the penalty terms is small.

In order to explain the method, we consider an example of a ray path scheme consisting of seven nodes: two end nodes and five internal nodes, as shown in Figure 13 of Part III. The DoF of the full gradient and Hessian are shown in Table 1. Each small cell is a vector of length 3. The external DoF, labeled $S$ and $R$, are shaded green, while the internal DoF, labeled $A$ and $B$, are shaded yellow. The cells labeled by $S$ and $R$ contain the locations of the source and receiver, respectively, $B$ contains the ray velocity direction at the receiver, and $A$ (a longer cell) contains all other spatial and directional DoF. The cells labeled by $S, R$ and $B$ are of length 3. There are $N+1$ nodes enumerated from zero to $N$ (i.e., $N$ intervals). The length of cell $A$ is $3(2N-1)$. The Hessian structure is presented in Tables 2a and 2b. Table 2a is the full scheme that includes all Hessian blocks. The diagonal blocks $SS, AA, RR$ and $BB$ in Table 2a of the global traveltime Hessian are square matrices. The bandwidth of the whole traveltime Hessian is 12 or 18, so that blocks $SR, SB$ and their corresponding transposed blocks may prove to be zero. Table 2b is a condensed scheme, where all DoF have been split into two groups: external (locations of the source and receiver) and internal (locations of internal nodes and directions of all nodes), labeled $E$ and $I$, respectively. The condensed scheme has been implemented in the suggested compression algorithm.

One should recall that the traveltime gradient vanishes only for the imposed boundary conditions (BC), where the locations of the two endpoints are fixed. However, when computing the geometric spreading, the endpoint locations are no longer fixed, so only the gradient components related to the internal DoF (other than endpoint locations) vanish. The corresponding gradient parts are,

$$\nabla_S t \neq 0 \;\;,\;\; \nabla_A t = 0 \;\;,\;\; \nabla_R t \neq 0 \;\;,\;\; \nabla_B t = 0 \qquad . \qquad \text{(A1)}$$

Now assume that $\Delta\mathbf{d}_S$ and $\Delta\mathbf{d}_R$ are small displacements wrt the locations of the end nodes and $R$ , respectively. This immediately means that all other locations and all the ray directions (including the directions at the endpoints) obtain increments $\Delta\mathbf{d}_A$ and $\Delta\mathbf{d}_B$, in order to achieve a new, slightly shifted, but still stationary ray path. The gradient of the external DoF changes, while the gradient of the internal DoF does not change – it has to remain zero for any new stationary path. This can be expressed by the block-matrix equation, in terms of the global Hessian and gradient, illustrated in Table 3. Cells with unknown values are shaded blue. Variations of the Hessian components can be ignored. Shifts $\Delta\mathbf{d}_S$ and $\Delta\mathbf{d}_R$ are infinitesimal; therefore, the change of the traveltime gradient can be linearized to find the variation between the new and old ray paths. The equation set corresponding to Table 3 reads,

$$\underbrace{\begin{bmatrix} \nabla_S\nabla_S t & \nabla_S\nabla_A t & \nabla_S\nabla_R t & \nabla_S\nabla_B t \\ \nabla_A\nabla_S t & \nabla_A\nabla_A t & \nabla_A\nabla_R t & \nabla_A\nabla_B t \\ \nabla_R\nabla_S t & \nabla_R\nabla_A t & \nabla_R\nabla_R t & \nabla_R\nabla_B t \\ \nabla_B\nabla_S t & \nabla_B\nabla_A t & \nabla_B\nabla_R t & \nabla_B\nabla_B t \end{bmatrix}}_{6(N+1)\times 6(N+1)} \underbrace{\begin{bmatrix} \Delta\mathbf{d}_S \\ \Delta\mathbf{d}_A \\ \Delta\mathbf{d}_R \\ \Delta\mathbf{d}_B \end{bmatrix}}_{6(N+1)} = \underbrace{\begin{bmatrix} \nabla_S t_{\text{new}} - \nabla_S t_{\text{old}} \\ 0 \\ \nabla_R t_{\text{new}} - \nabla_R t_{\text{old}} \\ 0 \end{bmatrix}}_{6(N+1)} \qquad . \qquad \text{(A2)}$$

This is a linear set with four unknown vectors: internal DoF and external gradients for the perturbed stationary path: $\Delta\mathbf{d}_A$, $\Delta\mathbf{d}_B$, $\nabla_S t_{\text{new}}$ and $\nabla_R t_{\text{new}}$. However, the equations for the zero and non-zero gradients are decoupled: The second and fourth equations of set A2 represent two equations with two unknown variables,

$$\underbrace{\begin{bmatrix} \nabla_A\nabla_A t & \nabla_A\nabla_B t \\ \nabla_B\nabla_A t & \nabla_B\nabla_B t \end{bmatrix}}_{\text{matrix } 6N\times 6N} \underbrace{\begin{bmatrix} \Delta\mathbf{d}_A \\ \Delta\mathbf{d}_B \end{bmatrix}}_{6N} = -\underbrace{\begin{bmatrix} \nabla_A\nabla_S t & \nabla_A\nabla_R t \\ \nabla_B\nabla_S t & \nabla_B\nabla_R t \end{bmatrix}}_{\text{matrix } 6N\times 6} \underbrace{\begin{bmatrix} \Delta\mathbf{d}_S \\ \Delta\mathbf{d}_R \end{bmatrix}}_{6} \quad , \tag{A3}$$

where the coefficient matrices on both sides are known (computed), and vector $\begin{bmatrix} \Delta\mathbf{d}_S & \Delta\mathbf{d}_R \end{bmatrix}$ of the endpoint shifts is assumed known. We will later show that this vector is not needed.

It is convenient to introduce the following notations,

$$\nabla_I\nabla_I t = \underbrace{\begin{bmatrix} \nabla_A\nabla_A t & \nabla_A\nabla_B t \\ \nabla_B\nabla_A t & \nabla_B\nabla_B t \end{bmatrix}}_{6N\times 6N} \quad , \tag{A4}$$

$$\nabla_I\nabla_E t = \underbrace{\begin{bmatrix} \nabla_A\nabla_S t & \nabla_A\nabla_R t \\ \nabla_B\nabla_S t & \nabla_B\nabla_R t \end{bmatrix}}_{\text{matrix } 6N\times 6} \quad , \tag{A5}$$

$$\Delta\mathbf{d}_I = \underbrace{\begin{bmatrix} \Delta\mathbf{d}_A \\ \Delta\mathbf{d}_B \end{bmatrix}}_{6N} , \qquad \Delta\mathbf{d}_E = \underbrace{\begin{bmatrix} \Delta\mathbf{d}_S \\ \Delta\mathbf{d}_R \end{bmatrix}}_{6} \quad . \tag{A6}$$

where $\nabla_I\nabla_I t$ is a part of the global traveltime Hessian, where both rows and columns are related to the internal DoF $\Delta\mathbf{d}_I$, and $\nabla_I\nabla_E t$ is the other part, whose rows are related to the internal

DoF $\Delta\mathbf{d}_I$, and columns – to the external DoF $\Delta\mathbf{d}_E$. Labels $I$ and $E$ mean internal and exernal, respectively. With these notations, equation A3 simplifies to,

$$\nabla_I \nabla_I t \cdot \Delta\mathbf{d}_I = -\nabla_I \nabla_E t \cdot \Delta\mathbf{d}_E \quad . \tag{A7}$$

The solution of equation A7 reads,

$$\Delta\mathbf{d}_I = -\left(\nabla_I \nabla_I t\right)^{-1} \cdot \nabla_I \nabla_E t \cdot \Delta\mathbf{d}_E \quad . \tag{A8}$$

This relationship, however, cannot be implemented because the matrix with only the internal DoF, $\nabla_I \nabla_I t$, to be inverted in equation A8, has a vanishing determinant. Furthermore, it has $2N$ zero eigenvalues (a zero eigenvalue per each Cartesian triplet of internal node location or its derivative wrt the arclength). Therefore, the constraints should be imposed before inverting the matrix. The internal node shifts $\Delta\mathbf{d}_I$ represent variation of the path of a paraxial ray. They may have a component tangent to the ray, and a component in the plane normal to the ray, both vs. the arclength of the central ray. The tangent component cannot be defined by the Hessian matrix, but it is also not needed for the computation of the geometric spreading, caustic detection and other dynamic characteristics. Therefore, we can assume that the tangent component vanishes: At the nodes, paraxial variations of the stationary path are normal to the central ray,

$$\mathbf{u}_i \cdot \mathbf{r}_i = 0 \quad , \tag{A9}$$

where $\mathbf{u}_i = \Delta\mathbf{x}_i$, and $\Delta\mathbf{d}_i = \left[\mathbf{u}_i \quad \dot{\mathbf{u}}_i\right]$, and dot means the derivative wrt the arclength of the central ray. Relationship A9 holds for non-nodal points as well. We enforce it for points in the vicinity of the nodes, and this means that its derivative yields one more equation, for directional DoF,

$$\dot{\mathbf{u}}_i \cdot \mathbf{r}_i + \mathbf{u}_i \cdot \dot{\mathbf{r}}_i = 0 \qquad . \tag{A10}$$

We emphasize that $\dot{\mathbf{u}}$ is not the direction of the paraxial ray, therefore, $\dot{\mathbf{u}} \cdot \dot{\mathbf{u}} \neq 1$. It is a derivative of the normal variation of the path wrt the arclength of the central ray (rather than wrt its own arclength). We further comment that the direction vectors of both central and paraxial rays are normalized,

$$\mathbf{r} \cdot \mathbf{r} = 1 \ , \quad (\mathbf{r} + \Delta\mathbf{r}) \cdot (\mathbf{r} + \Delta\mathbf{r}) = 1 \ \rightarrow \ \mathbf{r} \cdot \Delta\mathbf{r} = 0 \qquad , \tag{A11}$$

where the second-order infinitesimal value $\Delta\mathbf{r} \cdot \Delta\mathbf{r}$ has been ignored. In this study, we do not deal with $\Delta\mathbf{r}$, but with $\dot{\mathbf{u}}$.

Recall the order of internal DoF: first, the ray direction at the source, then locations and directions for all internal nodes, and eventually, the ray direction at the receiver. Equations A9 and A10 can be arranged in the following matrix form,

$$\underbrace{\begin{bmatrix} \mathbf{r}_o & & & & & & & \cdots & \\ & \mathbf{r}_1 & & & & & & \cdots & \\ & \dot{\mathbf{r}}_1 & \mathbf{r}_1 & & & & & \cdots & \\ & & & \mathbf{r}_2 & & & & \cdots & \\ & & & \dot{\mathbf{r}}_2 & \mathbf{r}_2 & & & \cdots & \\ & & & & & \mathbf{r}_3 & & \cdots & \\ & & & & & \dot{\mathbf{r}}_3 & \mathbf{r}_3 & \cdots & \\ \cdots & \cdots & \cdots & \cdots & \cdots & \cdots & \cdots & \cdots & \\ & & & & & & & & \mathbf{r}_N \end{bmatrix}}_{2N \times 6N} \cdot \underbrace{\begin{bmatrix} \dot{\mathbf{u}}_o \\ \mathbf{u}_1 \\ \dot{\mathbf{u}}_1 \\ \mathbf{u}_2 \\ \dot{\mathbf{u}}_2 \\ \mathbf{u}_3 \\ \dot{\mathbf{u}}_o \\ \cdots \\ \dot{\mathbf{u}}_N \end{bmatrix}}_{6N} = - \underbrace{\begin{bmatrix} \mathbf{r}_o \cdot \mathbf{u}_o \\ \\ \\ \\ \\ \\ \\ \\ \mathbf{r}_N \cdot \mathbf{u}_N \end{bmatrix}}_{2N} \qquad . \tag{A12}$$

We assume here that the external DoF $\mathbf{u}_o$ and $\mathbf{u}_N$ are also normal to the central ray. The empty spaces in the matrix and right-side vector of equation A12 are zeros. Direction and curvature at the nodes of the central ray, $\mathbf{r}_i$ and $\dot{\mathbf{r}}_i$, are row vectors of length 3. We use notation $\mathbf{C}_\perp$ for $2N \times 6N$ constraint matrix on the left side of equation A12. The column vector of length $6N$ on the left side includes all internal DoF, it is $\Delta\mathbf{d}_I$. The column vector on the right side has length $2N$ and can be presented as,

$$\underbrace{\begin{bmatrix} \mathbf{u}_o \cdot \mathbf{r}_o \\ \\ \vdots \\ \\ \mathbf{u}_N \cdot \mathbf{r}_N \end{bmatrix}}_{2N} = \underbrace{\begin{bmatrix} \mathbf{r}_o & \\ & \ddots \\ & & \mathbf{r}_N \end{bmatrix}}_{2N\times 6} \cdot \underset{6}{\begin{bmatrix} \mathbf{u}_o \\ \mathbf{u}_N \end{bmatrix}} \quad , \tag{A13}$$

where the vector of length 6 on the right side represents external DoF,

$$\begin{bmatrix} \mathbf{u}_o \\ \mathbf{u}_N \end{bmatrix} \equiv \begin{bmatrix} \Delta\mathbf{d}_S \\ \Delta\mathbf{d}_R \end{bmatrix} \equiv \Delta\mathbf{d}_I \quad . \tag{A14}$$

We use notation $\mathbf{R}$ for the matrix of dimension $2N \times 6$ on the left side of equation A13. Combining equations A12-A14, and using the mentioned notations, we obtain,

$$\mathbf{C}_\perp \cdot \Delta\mathbf{d}_I = -\mathbf{R} \cdot \Delta\mathbf{d}_E \quad . \tag{A15}$$

Next, we solve the under-defined linear equation set A7 ($6N$ equations with $6N$ unknowns, but not all equations are independent), subjected to the linear constraint set A15 (additional $2N$ equations with the same $6N$ unknowns), that makes it fully defined. We emphasize that the resulting combined set is neither under-defined nor over-defined. It is convenient to solve it formally with the least-squares fit, but due to the above statement, the result is exact (if we use Lagrangian multipliers instead of least squares, the result will be identical). For the same reason, the solution is insensitive to the weight of the constraint term, so we assume the weight is 1,

$$\left(\nabla_I \nabla_I t \cdot \Delta \mathbf{d}_I + \nabla_I \nabla_E t \cdot \Delta \mathbf{d}_E\right)^2 + \left(\mathbf{C}_\perp \cdot \Delta \mathbf{d}_I + \mathbf{R} \cdot \Delta \mathbf{d}_E\right)^2 \to \min \qquad , \qquad \text{(A16)}$$

where square means a scalar product of a vector by itself. Minimization of a scalar function in equation A16 is required wrt the unknown internal DoF $\Delta \mathbf{d}_I$ . Equation A16 leads to a linear set,

$$\left[\left(\nabla_I \nabla_I t\right)^T \cdot \nabla_I \nabla_I t + \mathbf{C}^T \cdot \mathbf{C}\right] \cdot \Delta \mathbf{d}_I = -\left[\left(\nabla_I \nabla_I t\right)^T \cdot \nabla_I \nabla_E t + \mathbf{C}^T \cdot \mathbf{R}\right] \cdot \Delta \mathbf{d}_E \qquad . \qquad \text{(A17)}$$

Now the matrix on the left side of equation A17 is invertible (it has a non-vanishing determinant). This matrix is narrow-band and symmetric. Note also that the traveltime Hessian, $\nabla_I \nabla_I t$ , related to the internal DoF, is a symmetric matrix. The solution of equation A17 reads,

$$\Delta \mathbf{d}_I = -\left[\left(\nabla_I \nabla_I t\right)^2 + \mathbf{C}_\perp^T \cdot \mathbf{C}_\perp\right]^{-1} \left(\nabla_I \nabla_I t \cdot \nabla_I \nabla_E t + \mathbf{C}_\perp^T \cdot \mathbf{R}\right) \cdot \Delta \mathbf{d}_E \qquad . \qquad \text{(A18)}$$

Solution A18 shows the variation of the internal DoF $\Delta \mathbf{d}_I$ vs. the variation of the external DoF $\Delta \mathbf{d}_E$ . Internal DoF are considered dependent in this analysis. They adjust in such a way that the internal gradient components remain zero for a perturbed stationary path. Actually, the above equation represents the normal nodal shifts of a paraxial ray (and their derivatives wrt the arclength of the central ray), whose trajectory is stationary as well.

Next, we consider the first and third equations of set A2,

$$\underbrace{\begin{bmatrix} \nabla_S \nabla_S t & \nabla_S \nabla_R t \\ \nabla_R \nabla_S t & \nabla_R \nabla_R t \end{bmatrix}}_{6\times 6} \underbrace{\begin{bmatrix} \Delta \mathbf{d}_S \\ \Delta \mathbf{d}_R \end{bmatrix}}_{6} + \underbrace{\begin{bmatrix} \nabla_S \nabla_A t & \nabla_S \nabla_B t \\ \nabla_R \nabla_A t & \nabla_R \nabla_B t \end{bmatrix}}_{6\times 6N} \underbrace{\begin{bmatrix} \Delta \mathbf{d}_A \\ \Delta \mathbf{d}_B \end{bmatrix}}_{6N} = \underbrace{\begin{bmatrix} \Delta \mathbf{p}_S \\ \Delta \mathbf{p}_R \end{bmatrix}}_{6} \quad , \tag{A19}$$

where the right side represents variations of external slowness vectors (traveltime gradients),

$$\begin{bmatrix} \Delta \mathbf{p}_S \\ \Delta \mathbf{p}_R \end{bmatrix} = \begin{bmatrix} \nabla_S t_{\text{new}} - \nabla_S t_{\text{old}} \\ \nabla_R t_{\text{new}} - \nabla_R t_{\text{old}} \end{bmatrix} \quad . \tag{A20}$$

It is suitable to introduce the following notations,

$$\nabla_E \nabla_E t = \underbrace{\begin{bmatrix} \nabla_S \nabla_S t & \nabla_S \nabla_R t \\ \nabla_R \nabla_S t & \nabla_R \nabla_R t \end{bmatrix}}_{6\times 6} \quad , \tag{A21}$$

and

$$\nabla_E \nabla_I t = \underbrace{\begin{bmatrix} \nabla_S \nabla_A t & \nabla_S \nabla_B t \\ \nabla_R \nabla_A t & \nabla_R \nabla_B t \end{bmatrix}}_{6\times 6N} \quad , \tag{A22}$$

where $\nabla_E \nabla_E t$ is a part of the global traveltime Hessian, where both rows and columns are related to the external DoF $\Delta \mathbf{d}_E$, and $\nabla_E \nabla_I t$ is the other part, whose rows are related to the external DoF $\Delta \mathbf{d}_E$, and columns – to the internal DoF $\Delta \mathbf{d}_I$. With this notation, equation A19 simplifies to,

$$\nabla_E \nabla_E t \cdot \Delta \mathbf{d}_E + \nabla_E \nabla_I t \cdot \Delta \mathbf{d}_I = \begin{bmatrix} \Delta \mathbf{p}_S \\ \Delta \mathbf{p}_R \end{bmatrix} \quad . \tag{A23}$$

The change in slowness at the external nodes (the source and receiver can also be presented as a product of the compressed source-receiver traveltime Hessian $\nabla_E \nabla_E t$ and the shifts of these nodes (variations of the source and receiver locations),

$$\underbrace{\nabla_E \nabla_E t}_{6\times 6} \underbrace{\begin{bmatrix} \Delta \mathbf{d}_S \\ \Delta \mathbf{d}_R \end{bmatrix}}_{6} = \underbrace{\begin{bmatrix} \Delta \mathbf{p}_S \\ \Delta \mathbf{p}_R \end{bmatrix}}_{6} \quad . \tag{A24}$$

Combining equations A23 and A24, we obtain,

$$\nabla_E \nabla_E t \cdot \Delta \mathbf{d}_E + \nabla_E \nabla_I t \cdot \Delta \mathbf{d}_I = \nabla_E \nabla_E t \cdot \Delta \mathbf{d}_E \quad , \quad \text{where} \quad \Delta \mathbf{d}_E = \begin{bmatrix} \Delta \mathbf{d}_S \\ \Delta \mathbf{d}_R \end{bmatrix} \quad . \tag{A25}$$

Next, we introduce the internal DoF $\Delta \mathbf{d}_I$ from equation A18 into A25,

$$\begin{gathered} \underbrace{\nabla_E \nabla_E t}_{6\times 6} \cdot \underbrace{\Delta \mathbf{d}_E}_{6} - \underbrace{\nabla_E \nabla_I t}_{6\times 6N} \cdot \underbrace{\left[ \underbrace{\left(\nabla_I \nabla_I t\right)^2}_{6N\times 6N} + \underbrace{\mathbf{C}_\perp^T}_{6N\times 2N} \cdot \underbrace{\mathbf{C}_\perp}_{2N\times 6N} \right]^{-1}}_{6N\times 6N} \cdot \\ \cdot \underbrace{\left( \underbrace{\nabla_I \nabla_I t}_{6N\times 6N} \cdot \underbrace{\nabla_I \nabla_E t}_{6N\times 6} + \underbrace{\mathbf{C}_\perp^T}_{6N\times 2N} \cdot \underbrace{\mathbf{R}}_{2N\times 6} \right)}_{6N\times 6} \cdot \underbrace{\Delta \mathbf{d}_E}_{6} = \underbrace{\nabla_E \nabla_E t}_{6\times 6} \cdot \underbrace{\Delta \mathbf{d}_E}_{6} \quad , \end{gathered} \tag{A26}$$

which can be arranged as,

$$\left\{ \nabla_E \nabla_E t - \nabla_E \nabla_E t + \nabla_E \nabla_I t \cdot \left[ \left(\nabla_I \nabla_I t\right)^2 + \mathbf{C}^T \cdot \mathbf{C} \right]^{-1} \cdot \left( \nabla_I \nabla_I t \cdot \nabla_I \nabla_E t + \mathbf{C}^T \cdot \mathbf{R} \right) \right\} \cdot \Delta \mathbf{d}_E = 0 \ , \tag{A27}$$

where $\nabla_E \nabla_E t$ is the compressed source-receiver traveltime Hessian, and $\nabla_E \nabla_E t$ is the external block of the global traveltime Hessian. Should the external DoF be a specific vector, one would say that it is an eigenvector of the matrix in the outer brackets, with a vanishing eigenvalue. However, the identity A27 holds for any external DoF $\Delta \mathbf{d}_E$. The only requirement is that all nodal shifts – internal and external – are normal to the central ray. However, even this requirement is important for the internal DoF, and can be relaxed for the external DoF. Therefore, the matrix in the outer brackets vanishes, and equation A27 yields the compressed source-receiver traveltime Hessian,

$$\nabla_E \nabla_E t = \nabla_E \nabla_E t - \nabla_E \nabla_I t \cdot \left[ \left( \nabla_I \nabla_I t \right)^2 + \mathbf{C}_\perp^T \cdot \mathbf{C}_\perp \right]^{-1} \cdot \left( \nabla_I \nabla_I t \cdot \nabla_I \nabla_E t + \mathbf{C}_\perp^T \cdot \mathbf{R} \right) \quad . \qquad \text{(A28)}$$

Equation A28 connects the endpoint traveltime Hessian on the left-hand side to components (blocks) of the global traveltime Hessian on the right-hand side. This equation includes inversion of the large narrow-band symmetric positive-definite Hessian matrix with all the internal DoF. The inversion can be done with the Cholesky $\mathbf{U}^T \mathbf{U}$ decomposition, where $\mathbf{U}$ and $\mathbf{U}^T$ are the upper and lower triangular matrices, respectively, symmetric to each other. The decomposition is performed only once, and the backward substitution then runs for each column of the inverse. The inverse of the band matrix is fully populated.

The compressed source-receiver traveltime Hessian, like any matrix of the second derivatives, should be symmetric. The result of operator A28 is formally asymmetric. However, in all numerical examples that we checked, the skew-symmetry counterpart proved to be negligible. This part can be considered as a minor, acceptable operational inaccuracy of the approach. We remove the skew-symmetric part by computing the average of the $6 \times 6$ matrix and its transposed,

$$\nabla_{SR}\nabla_{SR}\, t = \frac{\left(\nabla_E \nabla_E\, t\right)^T + \nabla_E \nabla_E\, t}{2} \quad , \qquad \text{(A29)}$$

where $\nabla_{SR}\nabla_{SR}\, t$ is the final symmetric compressed source-receiver (endpoint) traveltime Hessian. This operator also averages the computational errors. The endpoint traveltime Hessian of size $6 \times 6$ can be split into four blocks of size $3 \times 3$, as shown in equation 5.

$$\nabla_{SR}\nabla_{SR}\, t = \underbrace{\begin{bmatrix} \nabla_S \nabla_S\, t & \nabla_S \nabla_R\, t \\ \nabla_R \nabla_S\, t & \nabla_R \nabla_R\, t \end{bmatrix}}_{6\times 6} = \begin{bmatrix} \dfrac{\partial^2 t}{\partial \mathbf{x}_S^2} & \dfrac{\partial^2 t}{\partial \mathbf{x}_S \partial \mathbf{x}_R} \\ \dfrac{\partial^2 t}{\partial \mathbf{x}_R \partial \mathbf{x}_S} & \dfrac{\partial^2 t}{\partial \mathbf{x}_R^2} \end{bmatrix} \quad . \qquad \text{(A30)}$$

The tilde over the Hessian blocks is used here to distinguish them from the corresponding blocks of the global traveltime Hessian matrix; the latter includes all DoF.

The symmetric $6 \times 6$ endpoint Hessian $\nabla_{SR}\nabla_{SR}\, t$ consists of four $3 \times 3$ blocks, and we compute it from the global Hessian, applying equation A28. In order to compute the geometric spreading we need only the mixed derivatives in the lower left block $\nabla_R \nabla_S\, t$,

$$\nabla_R \nabla_S\, t = \frac{\partial^2 t}{\partial \mathbf{x}_R \partial \mathbf{x}_S} \quad . \qquad \text{(A31)}$$

Each component of this $3 \times 3$ block is a mixed derivative of the traveltime, where one location component belongs to the source $S$ and the other to the receiver $R$.

## APPENDIX B. GEOMETRIC SPREADING FROM ENDPOINT SPATIAL TRAVELTIME HESSIAN

In this appendix, we describe a known procedure for computing geometric spreading. We follow the method suggested initially by Goldin (1986) and then applied by Červený (2000), Ursin and Hokstad (2003), Xu et al. (2005), Stovas and Ursin (2009), Ivanov and Stovas (2017), Xu et al. (2018), among others.

Consider surfaces $\Sigma_S$ and $\Sigma_R$ associated with the source and receiver, respectively. Each of these is characterized by a normal direction, $\mathbf{n}_S$ and $\mathbf{n}_R$, respectively. Two local frames of reference are introduced, so that in each case the local axis $x_3$ coincides with the normal to the surface. Two other axes are normal to $x_3$ and to each other, but can still rotate around $x_3$ (spin angle). The spin angles do not affect the results and can be assumed zero. Thus, we deal with zenith angles $\theta_S^{\mathbf{n}}, \theta_R^{\mathbf{n}}$ and azimuths $\psi_S^{\mathbf{n}}, \psi_R^{\mathbf{n}}$ of the two normal directions specified in the global frame. The global-to-local rotation matrices are,

$$\mathbf{A}_{\mathrm{rot}} = \begin{bmatrix} +\cos\theta_{\mathrm{rot}}^{\mathbf{n}} \cos\psi_{\mathrm{rot}}^{\mathbf{n}} & +\cos\theta_{\mathrm{rot}}^{\mathbf{n}} \sin\psi_{\mathrm{rot}}^{\mathbf{n}} & -\sin\theta_{\mathrm{rot}}^{\mathbf{n}} \\ -\sin\psi_{\mathrm{rot}}^{\mathbf{n}} & +\cos\psi_{\mathrm{rot}}^{\mathbf{n}} & 0 \\ \sin\theta_{\mathrm{rot}}^{\mathbf{n}} \cos\psi_{\mathrm{rot}}^{\mathbf{n}} & +\sin\theta_{\mathrm{rot}}^{\mathbf{n}} \sin\psi_{\mathrm{rot}}^{\mathbf{n}} & +\cos\theta_{\mathrm{rot}}^{\mathbf{n}} \end{bmatrix} , \qquad \text{(B1)}$$

where subscript “rot” should be replaced by $S$ and $R$, accordingly. As mentioned, $\mathbf{A}_{\mathrm{rot}}$ is the global-to-local transformation, where each row represents components of a local axis in the global frame and each column represents components of a global axis in the local frame. We convert matrix $\nabla_R \nabla_S\, t$ defined in equation A31 to the two local frames simultaneously,

$$\left(\nabla_R \nabla_S\, t\right)^{\Sigma} = \mathbf{A}_S \left(\nabla_R \nabla_S\, t\right) \mathbf{A}_R^T \qquad . \tag{B2}$$

Superscript $\Sigma$ means that the Hessian is converted to the "two (endpoint) local frames". In a particular case where the local acquisition surfaces of both endpoints (at the source and receiver locations) are horizontal, the transform matrices are identity matrices, $\mathbf{A}_S = \mathbf{A}_R = \mathbf{I}$. Eventually, we remove the third row and column from the local mixed Hessian $\left(\nabla_R \nabla_S\, t\right)^{\Sigma}$, and it becomes a $2 \times 2$ matrix, $\mathbf{M}_{RS}^{\Sigma}$, whose components are highlighted in green in Table 4. Finally, the geometric spreading becomes,

$$L_{GS}\left(R,S\right) = \sqrt{\left| \frac{\cos\theta_{\text{ray},S}\, \cos\theta_{\text{ray},R}}{\det \mathbf{M}_{RS}^{\Sigma}} \right|} \qquad , \tag{B3}$$

where $\theta_{\text{ray},S}$ and $\theta_{\text{ray},R}$ are angles between the source and receiver ray velocity directions and the normal directions to their corresponding surfaces $\mathbf{n}_S$ and $\mathbf{n}_R$, respectively,

$$\cos\theta_{\text{ray},S} = \mathbf{n}_S \cdot \mathbf{r}_S \ , \qquad \cos\theta_{\text{ray},R} = \mathbf{n}_R \cdot \mathbf{r}_R \qquad . \tag{B4}$$

The ray velocity directions at the endpoints, $\mathbf{r}_S$ and $\mathbf{r}_R$, which belong to internal DoF, are computed with the Eigenray target function minimization.

Note that the (relative) geometric spreading is reciprocal, $L_{GS}\left(R,S\right) = L_{GS}\left(S,R\right)$. It has units $L^2/T$, where $L$ stands for distance and $T$ for time. We also apply the unitless normalized geometric spreading $L_{GS}/\sigma$, which can be used to estimate the plausibility of the solution, based on the amplitude loss (see equation 10 of this part, taking into account the normalized geometric spreading at the endpoint, and equation 30 of Part V, accounting for the normalized geometric spreading at all points along the path).

## REFERENCES

Alkhalifah, T., and I. Tsvankin, 1995, Velocity analysis for transversely isotropic media: Geophysics, **60**, no. 5, 1550-1566.

Mirko van der Baan, 2004, Processing of anisotropic data in the $\tau - p$ domain: I – Geometric spreading and moveout corrections: Geophysics, **69**, no. 3, 719-730.

Beylkin, G., 1985, Imaging of discontinuities in the inverse scattering problem by inversion of a casual generalized Radon transform: Journal of Mathematical Physics, **26**, 99-108.

Bleistein, N., 1987, On the imaging of reflectors in the earth: Geophysics, **52**, no. 7, 931-942.

Bleistein, N., J. Cohen, and J. Stockwell, 2001, Mathematics of Multidimensional Seismic Imaging, Migration, and Inversion, Springer Publishing Co., NY, ISBN 0-387-95061-3.

Castle, R., 1988, Shifted hyperbola and normal moveout: SEG International Exposition and 58th Annual Meeting, Expanded Abstract, 894-896.

Červený, V., 2000, Seismic ray theory: Cambridge University Press, ISBN 978-0521366717.

Chapman, C., 2004, Fundamentals of Seismic Wave Propagation: Cambridge University Press, ISBN 978-0521815583.

Fomel, S., and A. Stovas, 2010, Generalized nonhyperbolic moveout approximation: Geophysics, **75**, no. 2, U9-U18.

Gajewski, D., and I. Pšenčík, 1987, Computation of high-frequency seismic wavefields in 3D laterally inhomogeneous anisotropic media: Geophysical Journal of Royal Astronomical Society, **91**, 383-411.

Goldin, S., 1986, Seismic Traveltime Inversion, SEG, Tulsa, ISBN 978-0-93183-038-9.

Golikov, P., and A. Stovas, 2013, Moveout-based geometrical spreading approximation in TTI media: EAGE 75th Conference and Technical Exhibition, Expanded Abstract, DOI: 10.3997/2214-4609.20130617.

Gutenberg, B., 1936, The amplitudes of waves to be expected in seismic prospecting: Geophysics, **1**, no. 2, 252-256.

Hron, F., B. May, J. Covey, and P. Daley, 1986, Synthetic seismic sections for acoustic, elastic, anisotropic, and vertically inhomogeneous media: Geophysics, **51**, no. 3, 710-735.

Ivanov, Y., and A. Stovas, 2017, Mapping of geometrical spreading in anisotropic media: SEG International Exposition and 87th Annual Meeting, Expanded Abstract, 426-430.

Iversen E., B. Ursin, and M. de Hoop, 2018, High-order extrapolation of traveltime and geometrical spreading in anisotropic heterogeneous media: SEG International Exposition and 88th Annual Meeting, Expanded Abstract, 3383-3387.

Klimeš, L, 2006, Ray centred coordinate systems in anisotropic media: Studia Geophysica et Geodaetica, **50**, no. 3, 431-447.

Koren, Z., and I. Ravve, 2011, Full-azimuth subsurface angle domain wavefield decomposition and imaging: Part I – Directional and reflection image gathers: Geophysics, **76**, no. 1, S1-S13.

Koren, Z., and I. Ravve, 2019, Eigenray method: Geometric spreading: SEG International Exposition and 89th Annual Meeting, Expanded Abstract.

Najmi, A., 1996, Closed form solutions for the geometrical spreading in inhomogeneous media: Geophysics, **61**, no. 3, 1189-1197.

Newman, P., 1973, Divergence effects in a layered earth: Geophysics, **38**, no. 3, 481-488.

O'Doherty, R., and N. Anstey, 1971, Reflections on Amplitudes: Geophysical Prospecting, **19**, no. 3, 430-458.

Ravve, I., and Z. Koren, 2019, Eigenray method: Caustic detection: SEG International Exposition and 89th Annual Meeting, Expanded Abstract.

Schleicher, J., P. Hubral, and M. Tygel, 1993, Geometrical-spreading correction by a dual diffraction stack: Geophysics, **59,** no. 12, 1870-1873.

Schleicher, J., M. Tygel, B. Ursin, and N. Bleistein, 2001, The Kirchhoff-Helmholtz integral for anisotropic elastic media: Wave Motion, **34** (4), 353-364.

Schleicher, J., M. Tygel, and P., Hubral, 2007, Seismic true-amplitude imaging: Geophysical Developments no. **12**, ISBN 978-1560801436.

Shibo, X., A. Stovas, and Y. Sripanich, 2018, An anelliptic approximation for geometric spreading in transversely isotropic and orthorhombic media: Geophysics, **83**, no. 1, C37-C47.

Skopintseva, L., A. Aizenberg, M. Ayzenberg, M. Landrø, and T. Nefedkina, 2012, The effect of interface curvature on AVO inversion of near-critical and post-critical PP-reflections: Geophysics, **77**, no. 5, N1-N16.

Sommerfeld, A., 1964, Optics: Lectures on theoretical physics, 4: Academic Press, ISBN 978-0126546767.

Stovas, A., 2017, Geometrical spreading of P wave in acoustic orthorhombic media: SEG International Exposition and 87th Annual Meeting, Expanded Abstract, 400-404.

Stovas, A., 2018, Geometric spreading in orthorhombic media: Geophysics, **83**, no. 1, C61-C73.

Stovas, A. and B. Ursin, 2009, Improved geometric-spreading approximation in layered transversely isotropic media, Geophysics, **74**, no. 5, D85-D95.

Tanaka, T., H. Mikada, and J. Takekawa, 2017, S-wave AVO analysis for common scatter point gather for equivalent offset migration technique: SEG International Exposition and 87th Annual Meeting, Expanded Abstract, 723-727.

Tsvankin, I., and L. Thomsen, 1994, Nonhyperbolic reflection moveout in anisotropic media: Geophysics, **59**, no. 8, 1290-1304.

Tsvankin, I., and V. Grechka, 2011, Seismology of Azimuthally Anisotropic Media and Seismic Fracture Characterization, SEG, Oklahoma, ISBN 978-1-56080-228-0.

Tygel, M., J. Schleicher, and P. Hubral, 1992, Geometrical spreading corrections of offset reflections in a laterally inhomogeneous earth: Geophysics, **57**, no. 8, 1054-1063.

Ursin, B., 1990, Offset-dependent geometrical spreading in a layered medium: Geophysics, **55**, no. 4, 492-496.

Ursin, B., and K. Hokstad, 2003, Geometrical spreading in a layered transversely isotropic medium with vertical symmetry axis, Geophysics, **68**, no. 6, 2082-2091.

Vanelle, C., and D. Gajewski, 2002, Traveltime interpolation and geometrical spreading in 3-D anisotropic media: EAGE 64th Conference and Technical Exhibition, Expanded Abstract, 1-4.

Vanelle, C., and D. Gajewski, 2003, Determination of geometrical spreading from traveltimes: Journal of Applied Geophysics, **54**, no. 3, 391-400.

Vanelle, C., M. Spinner, T. Hertweck, C. Jäger, and D. Gajewski, 2006, Traveltime-based true-amplitude migration: Geophysics, **71**, no. 6, S251-S259.

Vanelle, C., and D. Gajewski, 2013, True-amplitude Kirchhoff depth migration in anisotropic media: The traveltime-based approach: Geophysics, **78**, no. 5, WC33-WC39.

Vasconcelos, I., and I. Tsvankin, 2006, Non-hyperbolic moveout inversion of wide-azimuth P-wave data for orthorhombic media: Geophysical Prospecting, **54**, no. 5, 535-552.

Virieux, J., V. Farra, and R. Madariaga, 1988, Ray tracing for earthquake location in laterally heterogeneous media: Journal of Geophysical Research, **93**, no. B6, 6585-6599.

Xu. S, and A. Stovas, 2018, Generalized non-hyperbolic approximation for qP-wave relative geometrical spreading in a layered transversely isotropic medium with a vertical symmetry axis: Geophysical Prospecting, **66**, 1290-1302.

Xu, S., A. Stovas, and Y. Sripanich, 2018, An anelliptic approximation for geometric spreading in transversely isotropic and orthorhombic media: Geophysics, **83**, no. 1, C37-C47.

Xu, X., I. Tsvankin, and A. Pech, 2005, Geometrical spreading of P-waves in horizontally layered, azimuthally anisotropic media, Geophysics, **70**, no. 5, D43-D53.

Xu, X., and I. Tsvankin, 2006, Anisotropic geometrical-spreading correction for wide-azimuth P-wave reflections: Geophysics, **71**, no. 3, D161-D170.

Xu, X., and I. Tsvankin, 2008, Moveout-based geometrical-spreading correction for PS-waves in layered anisotropic media: Journal of Geophysics and Engineering, **5**, no. 2, 195-202.

**LIST OF TABLES**

Table 1. External and internal DoF for seven-node assembly.

| node | DoF | Symbol | Property |
|---|---|---|---|
| 0 | $\mathbf{x}_0$ | $S$ | External |
| | $\mathbf{r}_0$ | $A$ | Internal |
| 1 | $\mathbf{x}_1$ | | |
| | $\mathbf{r}_1$ | | |
| 2 | $\mathbf{x}_2$ | | |
| | $\mathbf{x}_2$ | | |
| 3 | $\mathbf{x}_3$ | | |
| | $\mathbf{r}_3$ | | |
| 4 | $\mathbf{x}_4$ | | |
| | $\mathbf{r}_4$ | | |
| 5 | $\mathbf{x}_5$ | | |
| | $\mathbf{r}_5$ | | |
| 6 | $\mathbf{x}_6$ | $R$ | External |
| | $\mathbf{r}_6$ | $B$ | Internal |

Table 2. Blocks of traveltime Hessian related to external and internal DoF.

a) Full scheme

| | | | |
|---|---|---|---|
| $SS$ | $SA$ | $SR$ | $SB$ |
| $AS$ | $AA$ | $AR$ | $AB$ |
| $RS$ | $RA$ | $RR$ | $RB$ |
| $BS$ | $BA$ | $BR$ | $BB$ |

b) Condensed scheme ($E$ and $I$ are external and internal DoF, respectively)

| | |
|---|---|
| $EE$ | $EI$ |
| $IE$ | $II$ |

Table 3. Variation of traveltime gradient due to end node location change.

| | | | | | | | |
|---|---|---|---|---|---|---|---|
| $\nabla_S \nabla_S t$ | $\nabla_S \nabla_A t$ | $\nabla_S \nabla_R t$ | $\nabla_S \nabla_B t$ | | $\Delta \mathbf{d}_{\mathrm{S}}$ | | $\nabla_S t_{\mathrm{new}} - \nabla_S t_{\mathrm{old}}$ |
| $\nabla_A \nabla_S t$ | $\nabla_A \nabla_A t$ | $\nabla_A \nabla_R t$ | $\nabla_A \nabla_B t$ | $\cdot$ | $\Delta \mathbf{d}_{\mathrm{A}}$ | $=$ | 0 |
| $\nabla_R \nabla_S t$ | $\nabla_R \nabla_A t$ | $\nabla_R \nabla_R t$ | $\nabla_R \nabla_B t$ | | $\Delta \mathbf{d}_{\mathrm{R}}$ | | $\nabla_R t_{\mathrm{new}} - \nabla_R t_{\mathrm{old}}$ |
| $\nabla_B \nabla_S t$ | $\nabla_B \nabla_A t$ | $\nabla_B \nabla_R t$ | $\nabla_B \nabla_B t$ | | $\Delta \mathbf{d}_{\mathrm{B}}$ | | 0 |

Table 4. Final $2 \times 2$ mixed endpoint traveltime Hessian.

| | | |
|---|---|---|
| $M_{RS}^{\Sigma}(1,1)$ | $M_{RS}^{\Sigma}(1,2)$ | $M_{RS}^{\Sigma}(1,3)$ |
| $M_{RS}^{\Sigma}(2,1)$ | $M_{RS}^{\Sigma}(2,2)$ | $M_{RS}^{\Sigma}(2,3)$ |
| $M_{RS}^{\Sigma}(3,1)$ | $M_{RS}^{\Sigma}(3,2)$ | $M_{RS}^{\Sigma}(3,3)$ |